# $^{119}$Sn solid state NMR and Mössbauer spectroscopic studies of the intermediate-valent stannide CeRuSn

F. M. Schappacher[a], P. Khuntia[b], A.K. Rajarajan[b], M. Baenitz[b], J. A. Mydosh[b, c], B. Chevalier[d], S. F. Matar[b], R. Pöttgen[a]

[a] *Institut für Anorganische und Analytische Chemie, Universität Münster, Corrensstrasse 30, D-48149 Münster, Germany*

[b] *Max Planck Institut für Chemische Physik fester Stoffe, Nöthnitzer Strasse 40, D-01187 Dresden, Germany*

[c] *Kamerlingh Onnes Laboratory, Leiden University, 2300RA Leiden, The Netherlands*

[d] *CNRS, Université de Bordeaux, ICMCB, F-33608 Pessac Cedex, France*

The ternary stannide CeRuSn is a static mixed-valent cerium compound with an ordering of trivalent and intermediate-valent cerium on two distinct crystallographic sites. $^{119}$Sn Mössbauer spectra showed two electronically almost identical tin atoms at 323 K, while at 298 K and below (77 and 4.2 K) two tin sites can clearly be distinguished. $^{119}$Sn solid state NMR experiments are performed to probe the local hyperfine fields at the two different Sn sites. $^{119}$Sn NMR powder spectra are nicely fitted with two Sn sites with nearly the same magnetic anisotropy, but with different absolute shift values. Both Sn sites are strongly affected by crossover-like transitions between 100 and 280 K. This local-site study confirms the superstructure modulations found in previous investigations. Towards lower temperatures the powder spectra are broadened giving strong evidence for the antiferromagnetically ordered ground state.

## 1. Introduction

Ternary intermetallic cerium compounds $Ce_xRu_yX_z$ ($X$ = B, Al, Ga, In, Mg, Cd, Zn) exhibit unusual crystal chemical features. They often show extremely short Ce–Ru distances which are directly related to intermediate-valent or almost tetravalent cerium. This peculiar feature has first been observed for several $Ce_xRu_yIn_z$ intermetallics [1-4] and CeRuSn [5-8]. In the meantime also aluminium [9], gallium [10], magnesium [11], cadmium [12], and zinc [13] compounds with similar structural features (two or even multiple cerium sites) have been reported. Overviews on the crystal chemical data are given in [14] and [15]. Electronic structure calculations [6] show strong covalent Ce–Ru bonding for the intermediate-valent cerium sites in these compounds.

Although the structural prerequisites for the occurrence of intermediate or mixed cerium valence has been observed in more than 20 binary and ternary intermetallic compounds [14, 15], only few of them have been studied in detail with respect to their physical properties. Among the $Ce_xRu_yX_z$ intermetallics CeRuSn [5-8] shows the most complex behavior for the magnetic and transport properties. At room temperature CeRuSn adopts a superstructure of the monoclinic CeCoAl type with two crystallographyically independent cerium sites. The latter are ordered and can clearly be assigned to trivalent and intermediate-valent cerium. Below room temperature CeRuSn shows the formation of multiple incommensurate charge density wave modulations with a $q$ vector changing as a function of temperature [8]. This behavior is manifested in large thermal hysteretic effects in the magnetic susceptibility, in the specific heat, as well as in electronic and heat transport properties [7]. At 2.7 K long-range antiferromagnetism occurs with only half of the cerium sites participation [7].

In continuation of our structure-property investigations on CeRuSn we herein present a local-site spectroscopic investigation by means of temperature dependent $^{119}$Sn solid state NMR and $^{119}$Sn Mössbauer spectroscopy.

## 2. Experimental

*2.1. Synthesis*

Starting materials for the synthesis of the polycrystalline CeRuSn sample were cerium ingots (Sigma Aldrich), ruthenium powder (Degussa-Hüls, ca. 200 mesh), and tin granules (Merck), all with stated purities better than 99.9 %. The three elements were weighed in the ideal 1 : 1 : 1 atomic ratio and reacted by arc-melting [16] under an argon atmosphere of ca. 700 mbar. The argon was purified before with molecular sieves, silica gel, and titanium sponge (900 K). The product button was re-melted three times to ensure homogeneity. The total weight loss after the different re-melting procedures was smaller than 0.5 %. The resulting polycrystalline CeRuSn sample is stable in air over months. The phase purity of our sample was checked by a Guinier powder pattern (Cu Kα$_1$ radiation and α-quartz ($a$ = 491.30; $c$ = 540.46 pm) as internal standard, imaging plate technique (Fujifilm, BAS-READER 1800)) and by metallography in combination with EDX.

### 2.2. $^{119}$Sn Mössbauer spectroscopy

A Ca$^{119m}$SnO$_3$ source was available for the $^{119}$Sn Mössbauer spectroscopic investigation. The sample was placed within a thin-walled PMMA container at a thickness of about 10 mg Sn/cm$^2$. A palladium foil of 0.05 mm thickness was used to reduce the tin K X-rays concurrently emitted by this source. The measurement was conducted in the usual transmission geometry in the temperature range from 4.2 to 323 K with a total counting time of up to 2 days per spectrum.

### 2.3. $^{119}$Sn solid state NMR

Field swept $^{119}$Sn NMR measurements on polycrystalline CeRuSn were performed using a Tecmag spectrometer employing standard pulse techniques in the temperature range 1.8 ≤ T ≤ 290 K at 49.5 MHz. The spectral intensity was obtained by integrating the spin echo in the time domain. In order to determine the shift we measured SnO$_2$ powder as a non-magnetic reference with $^{119}$K = 0.5 % [17].

## 3. Results and Discussion

### 3.1. Crystal chemistry

The crystal structure of CeRuSn (space group $C2/m$, CeCoAl superstructure) has been discussed in detail in our original crystallographic work [5], together with a group-subgroup scheme explaining the superstructure formation. Herein we focus on the local coordination of the tin atoms which are the probes for $^{119}$Sn solid state NMR and $^{119}$Sn Mössbauer spectroscopy (*vide infra*).

A cutout of the CeRuSn structure is presented in Fig. 1 and the relevant interatomic distances for the two crystallographically independent tin atoms are listed in Table 1. The Ce1 atoms are in an intermediate-valent state (4–δ)+ while Ce2 is purely trivalent. This charge ordering has a drastic effect on the cerium coordination and upon superstructure formation. The cerium and ruthenium atoms show shifts of the atomic parameters, enabling the strong covalent Ce1–Ru bonding with distances of 233 and 246 pm, which are significantly shorter than the sum of the covalent radii of 289 pm. Longer Ce2–Ru distances of 288 and 291 pm are observed for the trivalent Ce2 atoms. The Ce1 and Ce2 atoms are stacked in a layer-like manner in the superstructure (Fig. 1).

The shifts of the cerium and ruthenium atoms then affect the tin coordination. In the superstructure both tin sites have site symmetry *m*, however, with different interatomic distances (Table 1). The most striking difference concerns the Sn–Sn distances. While Sn1–Sn2 of 309.8 pm is the shortest Sn–Sn distance in the structure (comparable to the structure of β-tin: 4 × 302 and 2 × 318 pm [18]), in the second tin coordination spherethe Sn1–Sn1 distance of 321.3 pm is distinctly shorter than Sn2–Sn2 of 338.5 pm. These differences affect the spectroscopic properties of the two tin sites.

## 3.2. $^{119}$Sn Mössbauer spectroscopy

The $^{119}$Sn Mössbauer spectra of CeRuSn at different temperatures are presented in Fig. 2 together with transmission integral fits. The corresponding fitting parameters are listed in Table 2. At 323 K, slightly above room temperature, the spectrum could be well reproduced with a single signal at an isomer shift of $\delta = 1.87(1)$ mm/s, subjected to quadrupole splitting of $\Delta E_Q = 1.47(1)$ mm/s. The isomer shift lies in the typical range for equiatomic and related stannides [19, 20]. The quadrupole splitting parameter reflects the non-cubic site symmetry of the tin atoms. The experimental line width parameter of 0.96(2) mm/s is slightly enhanced. We can therefore conclude that both tin sites are in a very close electronic state at 323 K (the temperature regime well above the phase transition observed in the susceptibility data [5]) and they cannot be distinguished by $^{119}$Sn Mössbauer spectroscopy. The tiny difference is only reflected in the slightly enhanced line width.

Already at 298 K the $^{119}$Sn spectrum is no longer symmetrical (Fig. 2). The spectra at 298, 77, and 4.2 K were best reproduced by a superposition of two signals of almost equal intensity, with different isomer shifts but similar quadrupole splitting parameters (Table 2). The isomer shifts $\delta_1 = 2.04$ and $\delta_2 = 1.80$ mm/s indicate higher *s* electron density at one of the tin nuclei. This is in line with the isomer shift systematics on a large series of tin compounds [21]. Again, for both signals we observe a slightly enhanced line width parameter. Keeping the modulation of the CeRuSn structure in going to lower temperature in mind, most likely these two signals are also just superpositions of electronically closely related tin atoms.

In order to assess the differences of electron populations especially on the two tin sites Sn1 and Sn2, we used full potential all electrons calculations within the augmented spherical wave ASW method. For all calculational details we refer to our previous manuscript [6]. At self consistent convergence of charges ($\Delta Q = 10^{-8}$) and energy ($\Delta Q = 10^{-8}$ eV) using a high integration of the Brillouin zone of the *C*-centered monoclinic Bravais lattice (2783 irreducible k-points generated from $20 \times 20 \times 20$ k-points), the calculation results imply a larger charge transfer on Sn1 versus Sn2, i.e. a more negatively charged Sn1 than Sn2 affecting the 5s and 5p subshells with the following occupations: Sn1 ($5s^{1.51}$, $5p^{2.22}$) and Sn2 ($5s^{1.41}$, $5p^{1.89}$), indicating that the $^{119}$Sn signal with the higher isomer shift value might correspond to Sn1. Such isomer shifts trends are known, e. g. for the series of Ca*T*Sn$_2$ (*T* = Rh, Pd, Ir) stannides [22].

These results are further illustrated by the site projected density of states (PDOS) for the spin-degenerate non-spin-polarized) configuration (Fig. 3). The use of a small scale along the *y*-axis is meant to show the Sn(*s* and *p*) different contributions between Sn1 (blue) and Sn2 (black dotted). The larger overall area below the PDOS is observed in both energy blocks relevant to the *s*-like PDOS, in the range {10,-7.5 eV} and the {-5 eV, $E_F$} range where Sn-*p* states mix with Ru *d*. Nevertheless the bonding will be expected mainly in the second energy region where Sn *p* is located.

At $E_F$ and above one can observe the 4*f* states of the two cerium sites: Ce1 whose *f* states are clearly above $E_F$ while Ce2 has its *f* states lying at $E_F$ with a large PDOS. This leads to a magnetic instability toward spin polarization affecting only Ce2 which carries a finite ordered magnetic moment [6].

*3.3. $^{119}$Sn solid state nuclear magnetic resonance*

Due to the presence of two distinct crystallographic different Sn sites with orthorhombic symmetry the $^{119}$Sn powder NMR spectra is rather broad and the fitting of the spectra is rather complex (Fig. 4). In Fig. 4 the $^{119}$Sn spin-echo intensity shows a noticeable broadening along with the appearance of a shoulder as the temperature is reduced below room temperature. The $^{119}$Sn NMR powder spectra could be consistently simulated by the superposition of two anisotropic lines (Sn' and Sn'') with comparable in-tensities (see Fig. 4 (b) and Fig. 4(c)). For simplification we assumed a tetragonal symmetry for the simulation leading to a 2:1 ratio in the intensities of the maxima of each individual line. The resulting shift values determined for the maximum position in the powder spectra (see arrow in Fig. 4 (b) and Fig. 4 (c)) for the Sn' and Sn'' sites are plotted as a function of temperature in Fig. 4 (d).

The overall T dependence of $^{119}$K(T) deviates strongly from the Curie –Weiss like behavior of the magnetic susceptibility (see Refs. 5 and 7). The positive shift is expected for cerium intermetallics from the conduction electron polarization model [17]. In general, the Knight shift has two components $K = K_0 + K_{4f}$. $K_0$ is the chemical shift which depends on the local environment and in most cases is negative and not T-dependent, but might change upon a structural phase transition, and $K_{4f}$ which is positive and reflects the polarization field transferred from the magnetic 4f-ion via conduction electron polarization. The absolute shift values for the Sn' line are larger than that of the Sn'' line which might be due to stronger polarization of the magnetic cerium. In the simplest approximation this results from a smaller distance to the site of the magnetic cerium and / or higher s-electron density (promoting the polarization effect). Upon cooling first the shift increases similar to the bulk susceptibility but towards lower temperatures the shift passes a maximum and seems to saturate at low temperatures (see Fig. 4). This is definitely caused by the broad structural transitions with onset at about 280 K. Here both, the chemical shift as well as the 4f contribution are both strongly affected. Unfortunately, it is difficult from the NMR to disentangle the two effects.

In general the situation and the complexity are reminiscent of CeRu$_4$Sn$_6$ with a tetragonal structure. Here also two anisotropic weakly shifted $^{119}$Sn NMR lines are observed [23]. In contrast to CeRu$_4$Sn$_6$ the equiatomic compound CeRuSn undergoes a crossover-like transition to a modulated superstructure [6-8] and the structure hosts two cerium species, one magnetic the other non-magnetic.

## 4. Conclusions

The unique equiatomic stannide CeRuSn has been characterized by $^{119}$Sn solid state NMR and Mössbauer spectroscopy. The superstructure formation, due to static mixed cerium valence, is clearly reflected in the low-temperature spectra. Two crystallographically independent tin sites are resolved by both spectroscopic techniques.

## Acknowledgments

This work was financially supported by the Deutsche Forschungsgemeinschaft.

## References

[1]  Zh. M. Kurenbaeva, A. I. Tursina, E. V. Murashova, S. N. Nesterenko, A. V. Gribanov, Yu. D. Seropegin, H. Noël, J. Alloys Compd. 442 (2007) 86-88.

[2]  E. V. Murashova, Zh. M. Kurenbaeva, A. I. Tursina, H. Noël, P. Rogl, A. V. Grytsiv, A. V. Gribanov, G. Giester, Yu. D. Seropegin, J. Alloys Compd. 442 (2007) 89-92.

[3]  A. I. Tursina, Zh. M. Kurenbaeva, A. V. Gribanov, H. Noël, T. Roisnel, Y. D. Seropegin, J. Alloys Compd. 442 (2007) 100-103.

[4]  E. V. Murashova, A. I. Tursina, Zh. M. Kurenbaeva, A. V. Gribanov, Yu. D. Seropegin, J. Alloys Compd. 454 (2008) 206-209.

[5]  J. F. Riecken, W. Hermes, B. Chevalier, R.-D. Hoffmann, F. M. Schappacher, R. Pöttgen, Z. Anorg. Allg. Chem. 633 (2007) 1094-1099.

[6]  S. F. Matar, J. F. Riecken, B. Chevalier, R. Pöttgen, V. Eyert, Phys. Rev. B 76 (2007) 174434-1–174434-6.

[7]  J. Mydosh, A. M. Strydom, M. Baenitz, B. Chevalier, W. Hermes, B. Chevalier, Phys. Rev. B 83 (2011) 054411-1–054411-6.

[8]  R. Feyerherm, E. Dudzik, S. Valencia, J. A. Mydosh, Y.-K. Huang, W. Hermes, R. Pöttgen, http://arxiv.org/abs/1111.5693v1, Phys. Rev. B, in press.

[9]  E. V. Murashova, A. I. Tursina, N. G. Bukhanko, S. N. Nesterenko, Zh. M. Kurenbaeva, Y. D. Seropegin, H. Noël, M. Potel, T. Roisnel, D. Kaczorowski, Mater. Res. Bull. 45 (2010) 993-999.

[10]  K. Shablinsaya, E. Murashova, A. Tursina, Z. Kurenbaeva, A. Yaroslavtsev, Y. Seropegin, Intermetallics, in press.

[11]  S. Linsinger, M. Eul, U. Ch. Rodewald, R. Pöttgen, Z. Naturforsch. 65b (2010) 1185-1190.

[12] F. Tappe, W. Hermes, M. Eul, R. Pöttgen, Intermetallics 17 (2009) 1035-1040.

[13] R. Mishra, W. Hermes, U. Ch. Rodewald, R.-D. Hoffmann, R. Pöttgen, Z. Anorg. Allg. Chem. 634 (2008) 470-474.

[14] W. Hermes, S. F. Matar, R. Pöttgen, Z. Naturforsch. 64b (2009) 901-908.

[15] T. Mishra, R.-D. Hoffmann, C. Schwickert, R. Pöttgen, Z. Naturforsch. 66b (2011) 771-776.

[16] R. Pöttgen, Th. Gulden, A. Simon, GIT Labor-Fachzeitschrift 43 (1999) 133-136.

[17] G. C. Carter, L. H. Bennett, D.J. Kahan, *Metallic Shifts in NMR*, Part 1 (Pergamonn Press) 1977.

[18] J. Donohue, The Structures of the Elements, Wiley, New York (U.S.A.) 1974.

[19] R. Mishra, R. Pöttgen, R.-D. Hoffmann, H. Trill, B. D. Mosel, H. Piotrowski, M. F. Zumdick, Z. Naturforsch. 56b (2001) 589-597.

[20] R.-D. Hoffmann, D. Kußmann, U. Ch. Rodewald, R. Pöttgen, C. Rosenhahn, B. D. Mosel, Z. Naturforsch. 54b (1999) 709-717.

[21] P. E. Lippens, Phys. Rev. B 60 (1999) 4576-4586.

[22] R.-D. Hoffmann, D. Kußmann, U. Ch. Rodewald, R. Pöttgen, C. Rosenhahn, B. D. Mosel, Z. Naturforsch. 54b (1999) 709-717.

[23] E. M. Brüning, M. Brando, M. Baenitz, A. Bentien, A. M. Strydom, R. E. Walstedt, F. Steglich, Phys. Rev. B 82 (2010) 125115.

Table 1

Interatomic distances (pm) for the two crystallographically independent tin atoms in the room temperature structure. The average distances are given in parentheses.

| Sn1: | 2 | Ru1 | 269.8 | Sn2: | 2 | Ru2 | 265.2 |
| --- | --- | --- | --- | --- | --- | --- | --- |
| | 1 | Ru2 | 273.9 | | 1 | Ru1 | 290.2 |
| | | | <271.2> | | | | <273.5> |
| | 1 | Sn2 | 309.8 | | 1 | Sn1 | 309.8 |
| | 1 | Sn1 | 321.3 | | 1 | Sn2 | 338.5 |
| | | | <315.6> | | | | <324.2> |
| | 2 | Ce1 | 338.5 | | 2 | Ce2 | 342.3 |
| | 2 | Ce2 | 347.1 | | 1 | Ce2 | 343.6 |
| | 1 | Ce1 | 353.4 | | 2 | Ce2 | 352.1 |
| | 2 | Ce1 | 360.9 | | 2 | Ce1 | 354.3 |
| | | | <349.5> | | | | <348.7> |

Table 2

Fitting parameters of $^{119}$Sn Mössbauer spectroscopic measurements of CeRuSn at different temperatures: $\delta$, isomer shift; $\Delta E_Q$, electric quadrupole splitting; $\Gamma$, experimental line width.

| Temp / K | $\delta_1$ / mms$^{-1}$ | $\Gamma_1$ / mms$^{-1}$ | $\Delta E_{Q1}$ / mms$^{-1}$ | $\delta_2$ / mms$^{-1}$ | $\Gamma_2$ / mms$^{-1}$ | $\Delta E_{Q2}$ / mms$^{-1}$ | Area 1:2 / % | $\chi^2$ |
| --- | --- | --- | --- | --- | --- | --- | --- | --- |
| 323 | 1.87(1) | 0.96(2) | 1.47(1) | | | | | 1.11 |
| 298 | 1.97(1) | 0.85(1) | 1.46(1) | 1.77(1) | 0.94(1) | 1.42(1) | 49:51(fix) | 1.42 |
| 77  | 1.99(1) | 0.86(2) | 1.49(1) | 1.80(2) | 1.01(1) | 1.39(1) | 49:51 | 2.10 |
| 4.2 | 2.04(1) | 0.91(1) | 1.52(1) | 1.80(1) | 1.07(2) | 1.48(1) | 49:51(fix) | 1.32 |

FIGURE CAPTIONS

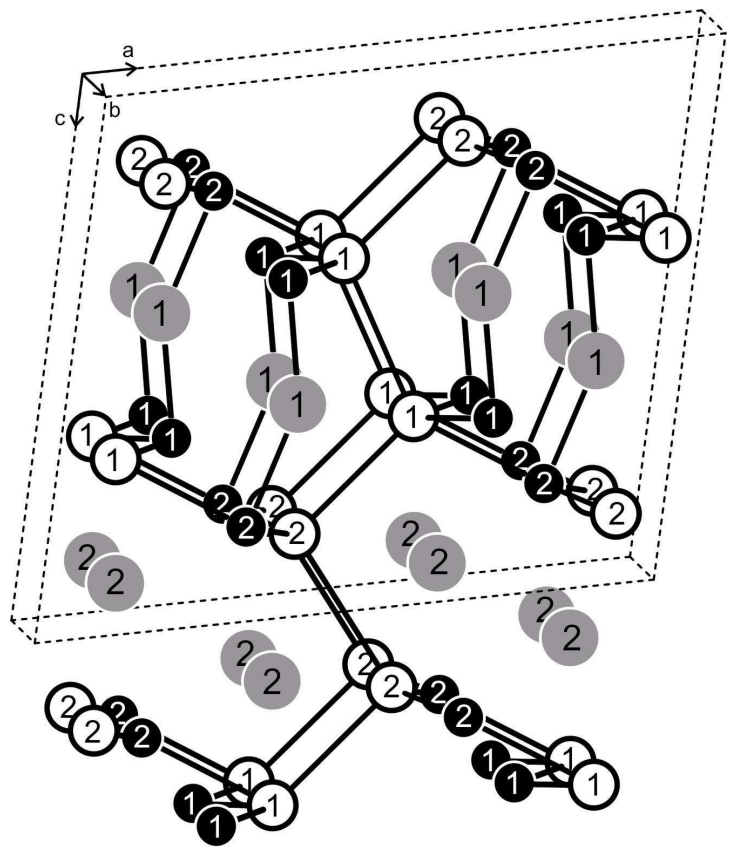

**Fig. 1.** Cutout of the CeRuSn structure. Cerium, ruthenium, and tin atoms are drawn as medium grey, black filled, and open circles, respectively. The shorter bonds are drawn. Atom designations are indicated. Ce1 is intermediate-valent; Ce2 trivalent.

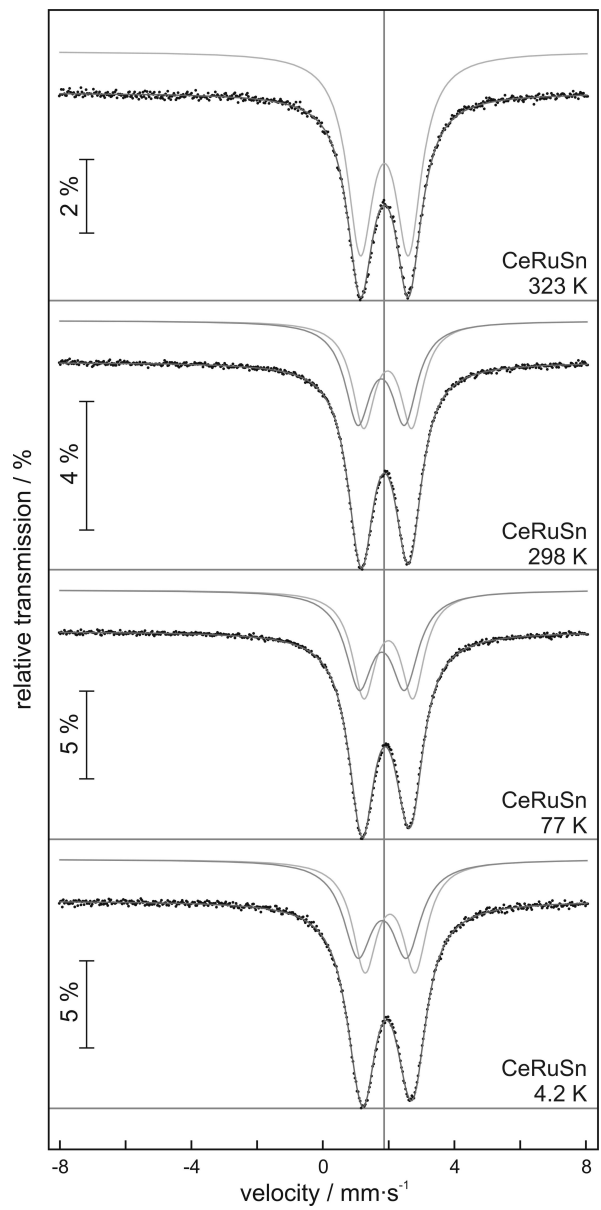

**Fig. 2.** $^{119}$Sn Mössbauer spectra of CeRuSn at various temperatures. The vertical line is introduced as a guide to the eye.

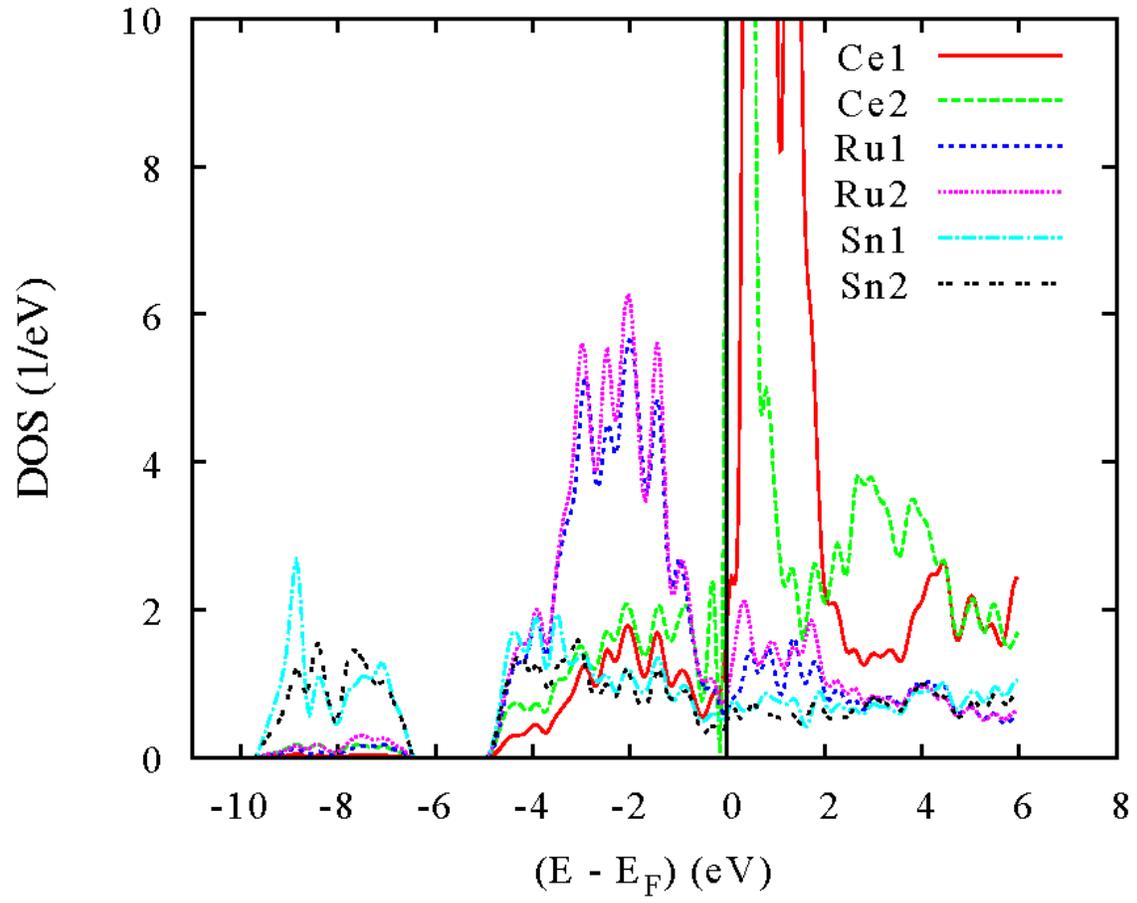

**Fig. 3 (color online).** Non-spin-polarized site-projected density of states (PDOS) of CeRuSn using a reduced DOS *y*-axis scale.

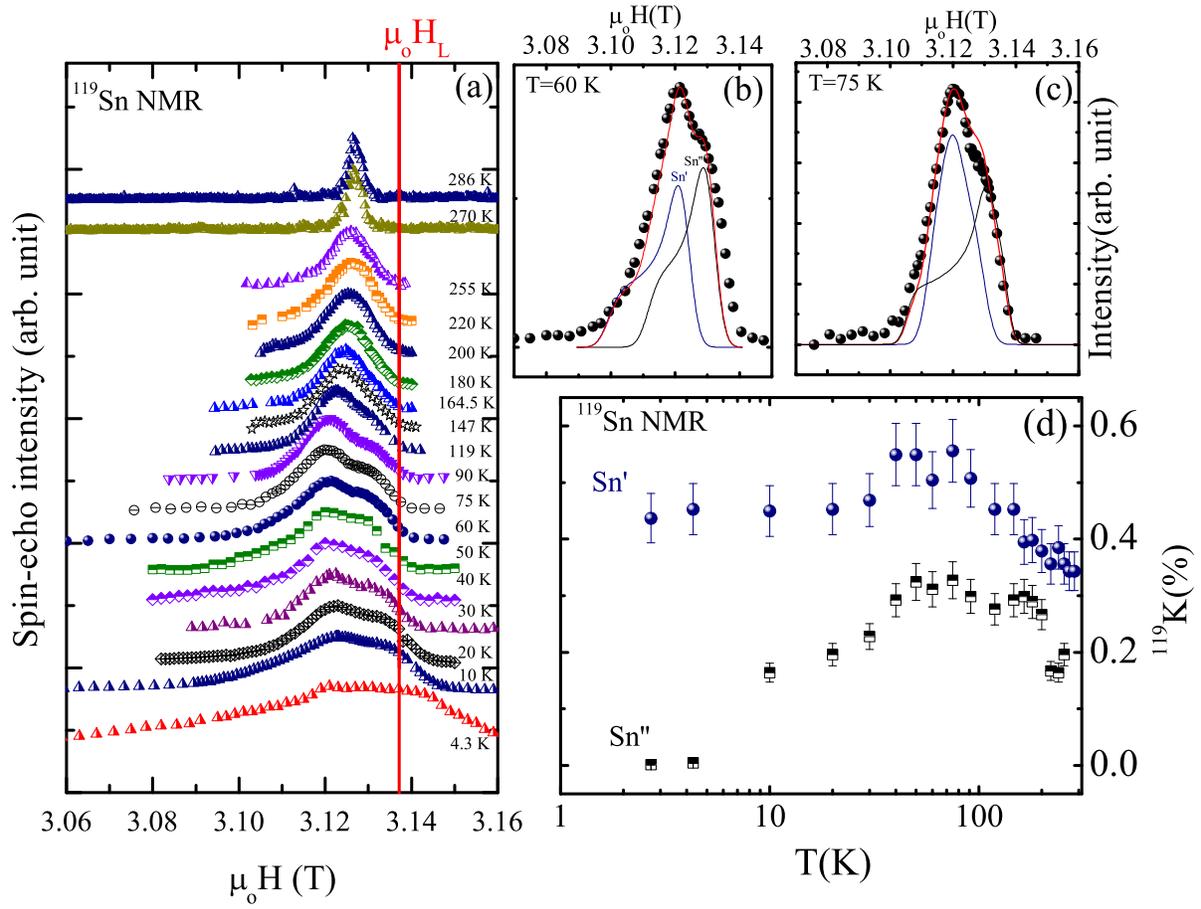

**Fig. 4 (color online).** (a) $^{119}$Sn powder NMR spectra at various temperatures taken at 49.5MHz. (b),(c) Comparison of the simulation with two Sn sites (Sn' and Sn'') and the experimentally obtained powder spectra at T = 60 K and at T = 75 K. (d) Temperature dependence of NMR shift for two $^{119}$Sn sites at 49.5 MHz.

13